\documentclass[article]{aa}  
\usepackage{graphicx}
\usepackage[multiple]{footmisc}
\usepackage{xcolor}

\usepackage{txfonts}
\usepackage{graphics, graphicx}
\usepackage{gensymb}
\newcommand{\Ha}{$\mathrm{H}\alpha$\xspace}
\newcommand{\Hb}{$\mathrm{H}\beta$\xspace}

\newcommand{\NIIb}{$[\mathrm{N}\textsc{ii}]\,\lambda 6584$\xspace}
\newcommand{\OIII}{$[\mathrm{O}\textsc{iii}]$\xspace}

\newcommand{\OIIIb}{$[\mathrm{O}\textsc{iii}]\,\lambda 5007$\xspace}

\newcommand{\LOIII}{$L_{[\mathrm{O}\textsc{iii}]}$\xspace}

\begin{document} 

       \title{The impact of environmental effects on AGN: a decline in the incidence of ionized outflows}

   \author{B.~Rodr\'iguez~Del~Pino\inst{1},         
          S.~Arribas\inst{1}, A. L. Chies-Santos\inst{2}, I. Lamperti\inst{1}, M. Perna\inst{1}, J. M. V\'ilchez\inst{4}. 
}
   \authorrunning{Rodr\'iguez Del Pino, et al. }

   \institute{Centro de Astrobiolog\'ia, (CAB, CSIC–INTA), Dpto. Astrof\'isica, Ctra. de Ajalvir Km. 4, 28850 Torrej\'on de Ardoz, Madrid, Spain\\
             \email:{brodriguez@cab.inta-csic.es}
   \and{Instituto de Física, Universidade Federal do Rio Grande do Sul (UFRGS), Av. Bento Gonçalves, 9500, Porto Alegre, RS, Brazil}
   \and{Instituto de Astrof\'isica de Andaluc\'ia. CSIC, Apartado de correos 3004, E-18080 Granada, Spain}
           }  

\date{Accepted}

  \abstract{
Active Galactic Nuclei (AGN) have been generally considered to be less frequent in denser environments due to the lower number of galaxy-galaxy interactions and/or the removal of their gas-rich reservoirs by the dense intergalactic medium. However, recent observational and theoretical works suggest that the effect of ram-pressure stripping acting on galaxies in dense environments might reduce the angular momentum of their gas, causing it to infall towards the super massive black hole (SMBH) at their centre, activating the AGN phase. In this work we explore the connection between environment and nuclear activity by evaluating the variation in the incidence of ionized outflows, a common phenomenon associated to nuclear activity, in AGN across different environments. We select a sample of $\sim3300$ optical AGN from the Sloan Digital Sky Survey Data Release 13 that we match with the group catalogue from Lim et al. 2017 to identify galaxies in isolation and residing in groups. We further probe their environment through the projected distance to the central galaxy of the group/cluster and the projected surface density to the 5th neighbour ($\delta_5$). The presence of ionized outflows is determined through the modelling of the \OIIIb emission line. We find that at lower masses ($<$~$10^{10.3}$M$_{\odot}$), the fraction of ionized outflows is significantly lower in satellite ($\sim7$\%) than in isolated ($\sim22$\%) AGN,  probably due to their different AGN luminosity, \LOIII, in that stellar mass range. The fraction of outflows decreases towards closer distances to the central galaxy of the group/cluster for all satellite AGN; however, only the lower-mass ones display a significant decline with $\delta_5$. Although this study does not include AGN in the densest regions of galaxy clusters, our findings suggest that AGN in dense environments accrete less gas than those in the field potentially due to the removal of the gas reservoirs via stripping or starvation, leading to a negative connection between environment and AGN activity. Based on our results, we propose that the observed change in the incidence of outflows towards denser regions of groups and clusters could contribute to the higher gas metallicities of cluster galaxies compared to field ones, especially at lower masses. 
}

   \keywords{AGN --
                outflows --
                kinematics and dynamics 
                clusters
               }

   \maketitle
%

\section{Introduction}
Galaxies evolve through cosmic time subject to the availability of gas reservoirs to fuel their stellar mass growth through star formation and the supermassive black holes (SMBH) at their centres. However, the availability and properties of the gas might be altered by processes taking place within the host galaxies or external ones acting on them. Internally, one of the processes that is considered very relevant on galaxy evolution is feedback from Active Galactic Nuclei (AGN) in the form of radiation, winds and jets that originate as a consequence of the feeding of the SMBH and can lead to the heating or ejection of the gas in the interstellar medium \citep[see][for a review]{fabianObservationalEvidenceActive2012}. This feedback from AGN is believed to play a crucial role in the evolution of galaxies by, for instance, preventing an excessive growth of galaxies through gas accretion that would lead to many more massive systems than observed \citep[][and references therein]{kormendyCoevolutionNotSupermassive2013}. Externally, galaxies can experience strong interactions between them via mergers \citep[][and references therein]{matzkoGalaxyPairsSloan2022} or tidal encounters \citep[][and references therein]{martigTriggeringMergerinducedStarbursts2008} that can lead to a more efficient consumption of the gas by triggering AGN activity and star formation. Additionally, galaxies residing in galaxy groups and clusters experience the interaction with the high-density surrounding media via processes such as the removal of the hot, diffuse gas when a small halo is accreted by a larger one  \citep[`strangulation', ][]{larsonEvolutionDiskGalaxies1980, vandenboschImportanceSatelliteQuenching2008} or the removal of the gas from the galaxy when the external pressure is sufficiently high \citep[ram-pressure stripping;][]{gunnInfallMatterClusters1972}. 

Given that environmental processes acting on galaxies can have a strong effect on their gas reservoirs, such processes might also be connected with the fuelling of AGN activity. Such connection has been investigated by several works, in general finding a lower fraction of AGN in galaxy clusters that was ascribed to less-frequent interactions between galaxies in denser regions that could trigger AGN activity \citep{gislerClustersGalaxiesStatistics1978, dresslerStatisticsEmissionlineGalaxies1985, popessoAGNFractionVelocity2006, pimbbletDriversAGNActivity2013, lopesNoSOCSSDSSVI2017b} or to a lack of a cold gas supply to feed the SMBH \citep{sabaterEffectInteractionsEnvironment2013}. However, other works have found that the AGN fraction remains roughly constant at different local densities \citep{millerEnvironmentActiveGalactic2003a, martiniSpectroscopicConfirmationLarge2006, rodriguezdelpinoOMEGAOSIRISMapping2017}. Interestingly, regions where AGN hosts reside tend to contain a higher fraction of blue, star-forming galaxies than regions around galaxies with no AGN \citep{coldwellRelationSeyfertAccretion2014}, indicating that AGN are prone to live in gas-rich environments. In such case, a lower fraction of AGN in galaxy groups and clusters could be explained by the lack of gas-rich regions due to environmental effects.  

The study of the connection between AGN activity and environment received new insights from the work by \citet{poggiantiRampressureFeedingSupermassive2017} where they found AGN activity in six out of seven massive galaxies ($4\times10^{10}-3\times10^{11}$M$_{\odot}$) undergoing extreme ram-pressure stripping (so-called jellyfish galaxies because of the striking tails of stripped gas that resemble the tentacles of the animal), a result that was interpreted as evidence that strong interactions with the intra-cluster medium (ICM) could trigger AGN activity. This interpretation was further supported by the fact that most of the galaxies presented relative velocities and projected distances within the cluster (i.e., positions in the cluster phase-space diagram) compatible with intense ram-pressure stripping. This work was extended by \citet{pelusoExploringAGNRamPressure2022} to a larger sample of 115 jellyfish galaxies, finding that they have 1.5 times higher probability to host an AGN than similar, non-jellyfish, star-forming galaxies, and that the incidence of AGN in jellyfish increases with stellar mass. However, as discussed in \citet{boselliRamPressureStripping2022}, the results from \citet{poggiantiRampressureFeedingSupermassive2017} are based on a very limited sample
of galaxies and, therefore, suffer from low-number statistics, whereas in \citet{pelusoExploringAGNRamPressure2022} the AGN sample also included LINERs (Low ionization Nuclear Emission-Line Regions) that are not always associated with nuclear activity \citep{singhNatureLINERGalaxies2013a, belfioreSDSSIVMaNGA2016b}. A different result was obtained in the rich population of 70 jellyfish galaxies in the multi-cluster system A901/2 at $z\sim0.165$, where \citet{roman-oliveiraOMEGAOSIRISMapping2019} found that only $7\%$ displayed signs of AGN activity. In their work, \citet{roman-oliveiraOMEGAOSIRISMapping2019} included jellyfish galaxies with different degrees of stripping and a wider range of stellar masses ($10^{9}-10^{11.5}$M$_{\odot}$), and identified AGN activity through the WHAN diagnostic diagram \citep{cidfernandesAlternativeDiagnosticDiagrams2010} instead of the BPT diagram \citep{baldwinClassificationParametersEmissionline1981a} used in the other works. Nevertheless, by applying the WHAN AGN selection criteria, \citet{roman-oliveiraOMEGAOSIRISMapping2019} found only a $\sim20\%$ AGN incidence in a sample of 42 jellyfish galaxies from the parent sample where the galaxies studied in \citet{poggiantiRampressureFeedingSupermassive2017} were taken from, i.e. the GASP survey \citep{poggiantiGASPGasStripping2017}.

In addition to be associated with galaxy clusters, recent studies have also found galaxies undergoing ram-pressure stripping in lower-density groups of halo masses $\sim10^{13}$M$_{\odot}$, although in these environments the frequency and strength of the jellyfish feature appears much smaller than in clusters \citep{robertsLoTSSJellyfishGalaxies2021a, kolcuQuantifyingRoleRam2022}. In the work by \citet{kolcuQuantifyingRoleRam2022} they found that only two of their 30 tentative jellyfish candidates and none of their 13 securely-classified jellyfish displayed optical line ratios compatible with AGN activity. Simulations of galaxies undergoing ram-pressure stripping have shown that, as a consequence of the interaction with the ICM, gas in galaxies can lose angular momentum and spiral down into the central parts of the galaxy \citep{schulzMultiStageThreedimensional2001, tonnesenGASSTRIPPINGSIMULATED2009, tonnesenStarFormationRam2012}. Some simulations predict that this gas can trigger AGN activity in galaxies with stellar masses above $10^{9.5}$M$_{\odot}$, whereas in less massive systems ram-pressure stripping suppresses both star formation and SMBH accretion \citep{marshallTriggeringActiveGalactic2018, ricarteLinkRamPressure2020}. Other theoretical works find that, when averaged over several galaxies and over long timescales, accretion onto the SMBH is significantly suppressed in clusters at all masses \citep[$10^{9.7}-10^{11.6}$M$_{\odot}$, ][]{joshiFateDiscGalaxies2020}.

In this work we explore the connection between environmental effects and AGN activity by evaluating the occurrence of feedback processes in the galaxies that originate from the feeding of the SMBH. In particular, we will focus on the study of galactic outflows since they are phenomena commonly observed in AGN, with fractions between $25\%$ and $40\%$ in optically-selected AGN, and are known to increase with galaxy properties such as the \OIIIb luminosity and stellar mass \citep{rodriguezdelpinoPropertiesionizedOutflows2019, wylezalekionizedGasOutflow2020}. The motivation behind this strategy arises from the fact that, if AGN activity can be triggered or hampered as a consequence of environmental effects disturbing the gas component in galaxies, a variation in the incidence of galactic outflows in AGN across different environments would also be an indication that environment is playing a role regulating AGN activity. In this regard, it is worth noting that galactic outflows also seem to be a common feature of AGN in jellyfish galaxies, as found in the more detailed analysis of the jellyfish presented in \citet{poggiantiRampressureFeedingSupermassive2017}, where four of the five confirmed AGN presented signatures of ionized outflows \citep{radovichGASPXIXAGN2019}. Interestingly, in one of these four sources,  \citet{georgeGASPXVIIIStar2019} found a cavity in the UV and CO~J$_{\rm 2-1}$ emission around the nucleus ($\sim9$~kpc) dominated by ionization from AGN. These results were interpreted as evidence for AGN feedback and environmental effects suppressing star formation in the galaxy. Although a possible connection between environmental effects and the occurrence of outflows in AGN has not yet been thoroughly explored in the literature, simulations such as those carried out by \citet{ricarteLinkRamPressure2020} have shown that AGN triggered by ram-pressure stripping produce observable outflows that can contribute to the quenching of star formation. More general studies such as those carried out by \citet{mcgeeOverconsumptionOutflowsQuenching2014} and \citet{trusslerBothStarvationOutflows2020} have suggested that the combination of these two mechanisms might explain the quenching of star formation in galaxies at different epochs, highlighting the fact that they operate on different timescales and their relevance has probably changed as a function of redshift. 

Finally, given that galactic outflows are considered to play a significant role in the regulation of metals in galaxies \citep{dayalPhysicsFundamentalMetallicity2013, arribasionizedGasOutflows2014a, rodriguezdelpinoPropertiesionizedOutflows2019} a change in the incidence of outflows in AGN with environment could also lead to differences in the metallicities of their host galaxies. In fact, galaxies residing in the field have been found to have lower metallicities for a given stellar mass compared to those residing in clusters \citep{cooperRoleEnvironmentMassmetallicity2008, ellisonMassMetallicityRelation2009} and in filaments \citep{aracilHighmetallicityPhotoionizedGas2006}. These differences have been generally ascribed to galaxies in low-density environments being more gas-rich and more metal-poor \citep{wuDependenceMassmetallicityRelation2017}, or to the accretion of metal-rich gas by satellite galaxies \citep{schaeferSDSSIVMaNGAEvidence2019}. In addition to these possible scenarios, a lower fraction of galactic outflows towards denser regions could prevent the ejection of metal-rich gas away from the galaxies and the subsequent reduction in global metallicity. 

With the aim of shedding more light on our understanding of the connection between AGN activity and environment, and its impact on galaxy evolution, in this work we investigate whether the incidence of ionized outflows in AGN varies depending on the environment where they reside. We make use of a large sample of AGN selected from the Sloan Digital Sky Survey (SDSS) Data Release 13 \citep[DR13; ][]{albareti13thDataRelease2017} in combination with the group catalogue from \citet{limGalaxyGroupsLowredshift2017}. We note that although the SDSS survey does not cover the more crowded regions of galaxy clusters due to fibre collisions \citep[e.g.,][]{gavazziCompleteCensusOptically2011}, it includes galaxies in groups and clusters where environmental effects have been shown to be at play \citep[e.g.,][]{baloghGalaxyEcologyGroups2004, pimbbletDriversAGNActivity2013}, including the identification of jellyfish galaxies \citep{robertsLoTSSJellyfishGalaxies2021a, kolcuQuantifyingRoleRam2022}.

In Section 2 we explain the selection of the AGN sample from SDSS DR13 and the definition of environmental parameters; in Section 3 we describe the spectral identification of ionized outflows; Section 4 contains  the study of the incidence of ionized outflows in galaxies residing in different environments; Section 5 includes a discussion of the results and in Section 6 we present the summary and main conclusions of this work. Throughout this work, we adopt a cosmology with $H_{\rm 0}=67.3$~km~s$^{-1}$ Mpc$^{-1}$, $\Omega_{\rm M}=0.315$ and $\Omega_{\Lambda}=0.685$ \citep{planckcollaborationPlanck2013Results2014a}. The $1\sigma$ binomial uncertainties for the fractions presented in this paper are calculated following \citet{cameronEstimationConfidenceIntervals2011}.

\section{Data}
\label{sec:data}

\subsection{Parent SDSS DR13}
\label{subsec:SDSSdata}

For this work we use the spectroscopic data from the SDSS Data Release 13 \citep{albareti13thDataRelease2017}, which is built upon previous releases but with significant improvements such as a better photometric calibration and measured redshifts for some of the fibre-collision galaxies. The survey is complete to an extinction-corrected Petrosian magnitude of 17.77~mag in the $r$ band.  In addition to the spectroscopic data, we use the set of MPA-JHU Value Added Catalogues\footnote{https://www.sdss.org/dr13/spectro/galaxy$\_$mpajhu}\footnote{http://www.mpa-garching.mpg.de/SDSS} that provide the main properties of the galaxies derived from the analysis of the SDSS photometric and spectroscopic data based on the methods of \citet{kauffmannStellarMassesStar2003}, \citet{brinchmannPhysicalPropertiesStarforming2004} and \citet{tremontiOriginMassMetallicityRelation2004a} such as: redshifts, emission-line fluxes, equivalent widths and total stellar masses. The provided emission line measurements were obtained after a careful modelling and subtraction of the stellar continuum. Stellar masses were derived from fitting of SDSS photometry as described in \citet{kauffmannStellarMassesStar2003}. In our work we only include galaxies with reliable line measurements and physical parameters (`RELIABLE'$=1$ in the MPA-JHU catalogues). In order to reduce the effects from redshift incompleteness we restrict our analysis to galaxies with stellar masses $>10^{9}$~M$_{\odot}$ and $z<0.08$, the redshift at which the number of galaxies in the selected stellar mass range starts to decrease. The physical sizes encompassed by the 3-arcsec-diametre SDSS fibres range from $0.6$~kpc at $z\sim0.01$ to $4.7$~kpc at $z\sim0.08$.

\subsection{AGN Selection}
\label{subsec:AGN_selection}

The selection of galaxies hosting an AGN is done using the standard optical line diagnostic BPT diagram \citep{baldwinClassificationParametersEmissionline1981a} that employs the ratios between the \OIIIb/\Hb and \NIIb/\Ha emission lines. Considering only the spectra where these four emission lines are detected with a signal-to-noise ratio ($SNR$) > 3, we select as AGN hosts the galaxies whose line ratios correspond to regions in the diagnostic diagram above the demarcation line from \citet{kewleyTheoreticalModelingStarburst2001b}, where theoretical models require the presence of nuclear activity to explain the ionization state of the gas. Additionally, we select only strong AGN cases using the separation line between Seyfert and LINERs suggested by \citet{cidfernandesAlternativeDiagnosticDiagrams2010} and require a minimum EW(\Ha) of 1.5~\AA{}. The application of these criteria implies that the \OIIIb line is detected with $SNR\gtrapprox10$ and leads to a sample of $\sim3300$ optically-selected AGN. 

We note that this selection includes both Type 1 and Type 2 AGN. However, since the line fluxes provided by the MPA-JHU Value Added Catalogues correspond to measurements using a single kinematic component, some Type 1 AGN might be missed by our selection since the emission from the Broad Line Region might lead to higher \Hb and \Ha fluxes and, as a consequence, to lower values of the \NIIb]/\Ha and \OIIIb/\Hb line ratios, placing them below the \citet{kewleyTheoreticalModelingStarburst2001b} demarcation line. However, we note that this effect does not lead to contamination in our selected sample from non-AGN galaxies, keeping our AGN selection pure, although somehow not complete.

\subsection{Environmental parameters}
\label{subsec:environment_parameters}
To probe the environment where AGN reside we use the group catalogue from \citet{limGalaxyGroupsLowredshift2017} that provides the results from applying to the SDSS DR13 data a halo-based group finder built upon those developed in \citet{yangHalobasedGalaxyGroup2005, yangGalaxyGroupsSDSS2007} and \citet{luGALAXYGROUPS2MASS2016} but with an improved halo mass assignment. From this catalogue we make use of the classification of galaxies into centrals (corresponding to either isolated galaxies or the most massive member of a group) and other group members, the number of galaxies in a group and the halo masses of groups assigned using abundance matching. We use halo mass estimates that consider galaxy stellar mass as a proxy and include only galaxy groups with halo masses that are complete at their corresponding redshifts. 

We further explore environmental effects using two parameters: (1) the projected surface density to the 5th neighbour around each AGN, $\delta_5$, considering only galaxies within $\pm$$\Delta$z~c~=~1000km~s$^{-1}$ and (2), for satellites, the projected distance to the central galaxy, $r_{\rm cen}$, normalized by $r_{\rm 180}$, the radius of the halo within which the mean mass density is 180 times the mean density of the Universe at the given redshift, obtained using Equation 4 from \citet{limGalaxyGroupsLowredshift2017}. Both parameters have been widely used to trace environmental changes in galaxies, with $\delta_5$ probing the local environment around the AGN \citep[e.g.,][]{dresslerGalaxyMorphologyRich1980, baloghGalaxyEcologyGroups2004, croomSAMIGalaxySurvey2021} and $r_{\rm cen}$ providing information on the influence of the intra-group or intra-cluster medium (which become denser at shorter distances) and the time since the galaxy enter the group environment \citep[e.g.,][]{bamfordGalaxyZooDependence2009, jaffeEffectEnvironmentGas2011}. 

Since we want to focus only on the effects that environment might have on the incidence of ionized outflows we need to avoid including galaxies that might be interacting with other objects via mergers. To this end, we use the visual morphological classification of SDSS galaxies carried out by the Galaxy Zoo project \citep{lintottGalaxyZooMorphologies2008, lintottGalaxyZooData2011a} that provide an estimate, based on the fraction of votes assigned by the classifiers, of whether a galaxy might be associated to a merger event. \citet{dargGalaxyZooFraction2010} performed a refined merger classification based on the Galaxy Zoo results for galaxies with redshifts $0.005 < z < 0.1$, finding that cases with a $weighted$-$merger$-$vote$ $fraction$ ($f_{\rm m}$)~$\gtrsim$~0.6 are generally robust mergers, whereas misclassifications start to be common for $f_{\rm m}$~$\lesssim$~$0.4$. To make a conservative removal of merger candidates, we discard from our sample all galaxies classified as mergers by \citet{dargGalaxyZooFraction2010} and for those outside their classified redshift range, we discard the galaxies with $f_{\rm m}$~$\geq$~$0.4$. In total we remove from our study 46 potential mergers (less than $1.5\%$ of the sample).

\begin{figure}
    \includegraphics[width=0.45\textwidth]{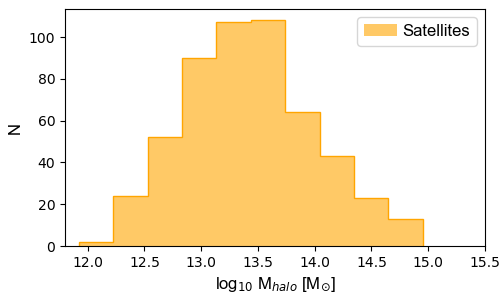}
    \caption[Histogram group halo masses]{Distribution of group halo masses for the selected sample of satellite AGN (see Section~\ref{subsec:environment_parameters}). } 
    \label{fig:Hist_halo_mass_sats}
\end{figure}

To work with a sample of AGN that are likely undergoing environmental effects we define a `satellite' sample containing the AGN that reside in groups of at least 5 members not being the central galaxy \citep[as in e.g.,][]{robertsLoTSSJellyfishGalaxies2021a}. Additionally, we define the `isolated' sample comprising the AGN with no group assignment that reside in regions where $\delta_5$ is within the lowest quintile among them. The central galaxies of the groups are not included in the satellite sample. Being the most massive ones, central galaxies are not expected to undergo strong interactions with the dense intra-group or intra-cluster medium. The satellite and isolated samples contain $\sim500$ and $\sim350$ AGN, respectively. The distribution of halo masses for the selected population of satellite AGN is shown in Figure~\ref{fig:Hist_halo_mass_sats}, ranging from $\sim10^{12}$ to $10^{15}$~M$_{\odot}$, with more than 75\% of them residing in groups of halo masses $>10^{13}$~M$_{\odot}$. Note that the remaining $\sim2500$ AGN not fulfilling our isolated and satellite definitions are still considered in this study as part of the full AGN sample.

\section{Modelling of the gas kinematics and outflow identification}
\label{sec:analysis}

To evaluate the presence of ionized outflows in our AGN sample we use the \OIIIb emission line because, being a forbidden line, its velocity profile is not affected by the high-velocities associated to the dense broad-line region in type 1 AGN. Instead, it traces the low-density ionized gas from the narrow-line region around the AGN and from the galaxy. Thus, a broadening of the \OIIIb emission would necessarily be a consequence of the presence of an additional ionized gas component moving at velocities different to the systemic one, allowing to detect, for instance, galactic outflows. Furthermore, \OIIIb is generally less affected by nearby emission or absorption lines that could hamper its spectral modelling (as happens with \Ha and the neighbouring {$[\mathrm{N}\textsc{ii}]\,\lambda\lambda6548,6584$ lines)\footnote{Although there is a HeI line at 5016\AA{} and Type 1 AGN might be contaminated from FeII lines and \Hb broad-line-region wings.}.

Before modelling the \OIIIb emission line we subtract the contribution of the stellar populations  from the spectra of the galaxies. The stellar continuum fitting is performed with the software Penalized Pixel-Fitting \citep[\textsc{pPXF};][]{cappellariImprovingFullSpectrum2017b} using as templates the \textsc{PEGASE-HR} simple stellar population (SSP) models \citep{leborgneEvolutionarySynthesisGalaxies2004b} that have the appropriate spectral resolution (FWHM~$\sim0.5$~\AA) and wavelength coverage (3900–6800~\AA) to fit the SDSS spectra. 

To reproduce the velocity distribution of the ionized gas we perform separate fits of the \OIIIb emission line using two different models with one and two kinematic components (narrow and broad), respectively, that are described as Gaussian functions. We remove from our sample the spectra ($<4\%$) affected by bad pixels within $\pm5\sigma$ of the narrow or broad components of the \OIIIb line. The two-component model is considered a better representation of the data based on the goodness of the fit, evaluated through the reduced-$\chi^2$ and the Bayesian Information Criterion statistics \citep{schwarzEstimatingDimensionModel1978}. We consider that an AGN hosts a candidate ionized outflow when the spectral fitting favours a model with a kinematic component (either the one from the single-component fit or the broad component in the two-component fit) with:

\begin{itemize}
\item a peak of emission at least 3 times larger than the standard deviation in the neighbouring continuum, 
\item a velocity dispersion (after deconvolution with the SDSS spectral line-spread function) larger than 200~kms$^{-1}$.
\end{itemize}

\begin{figure}
\begin{center}
    \includegraphics[width=0.49\textwidth]{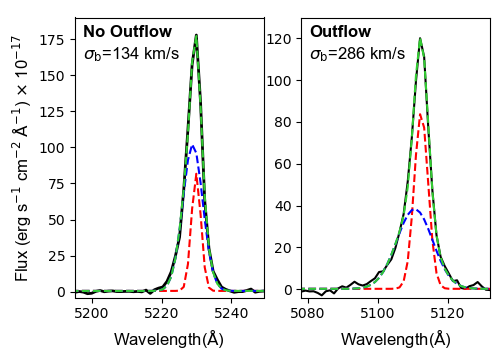}
    \caption[spectra outflow no outflow]{Example of the modelling of the gas kinematics through the \OIIIb line for two AGN of our sample. In both spectra a broad component (blue) is identified in addition to a narrow one (red). However, only the right one, with $\sigma_{\rm broad}>200$~kms$^{-1}$, is considered as an ionized outflow following our identification criteria (see Section~\ref{sec:analysis}). } 
    \label{fig:spectra_outflow_no_outflow}
\end{center}
\end{figure}

The adopted value of 200~kms$^{-1}$ is at least twice the systemic velocity dispersion measured throughout local galaxies within the stellar mass range we cover $\sim10^9-10^{\rm11.5}$M$_{\odot}$ \citep[e.g.,][]{greenHighStarFormation2010, yuWhatDrivesVelocity2019}, indicating that we are selecting clear cases where the galaxy hosts a secondary gas component following non-gravitational motions, potentially associated to a galactic outflow. A similar kinematic criterion has been widely used to identify ionized outflows in other works \citep[e.g.,][]{hoSAMIGalaxySurvey2014, wooPrevalenceGasOutflows2016, gallagherWidespreadStarFormation2019, forsterschreiberKMOS3DSurveyDemographics2019a, rodriguezdelpinoPropertiesionizedOutflows2019}. As an example, Figure~\ref{fig:spectra_outflow_no_outflow} contains the spectra of two galaxies where a model with two-kinematic components is preferred but only one of them hosts an ionized outflow following our selection criteria.

\begin{figure*}
    \centering
    \begin{minipage}{.99\textwidth}
    \centering
    \includegraphics[width=0.45\textwidth]{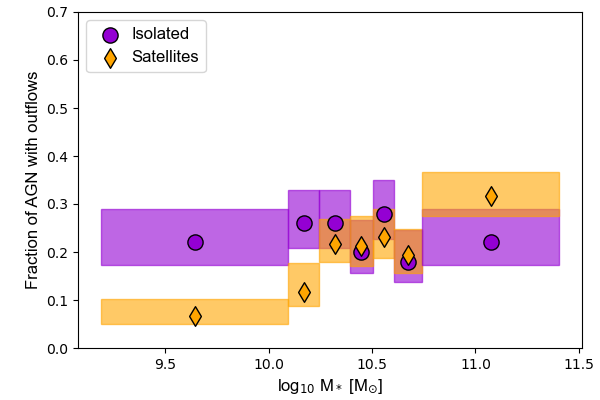}
    \includegraphics[width=0.45\textwidth]{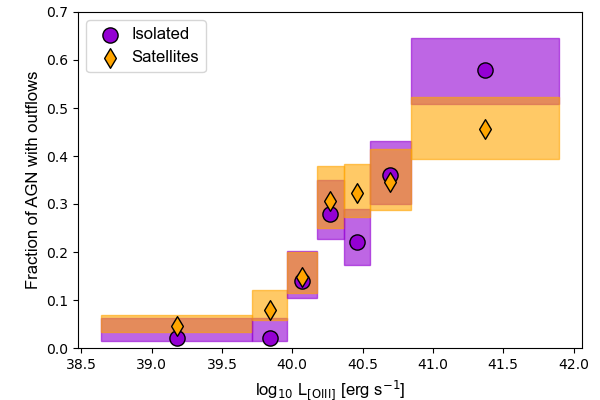}
    \caption[Fraction outflows stellar mass cens and sats]{Fraction of outflows as a function of stellar mass ($\emph{left}$) and \LOIII ($\emph{right}$) for isolated and satellite AGN. Heights of the boxes denote the 1$\sigma$ errors associated to the fractions. Bins are defined to include at least 50 galaxies each.} 
    \label{fig:Fraction_vs_mass_lum_cens_sats}
    \end{minipage}
\end{figure*}

\section{Incidence of ionized outflows in AGN as a function of environment}
\label{sec:incidence_outflows}

\subsection{Outflow incidence}
\label{subsec:outflow_incidence}

By applying the selection criteria described in the previous section to our sample of AGN we find that $24\%$ of them ($\sim800$) host an ionized outflow. This global incidence value is very similar to the $25\%$ reported by \citet{wylezalekionizedGasOutflow2020} in the MaNGA-selected sample of local AGN, where they also used the \OIIIb to trace outflows. However, our value is lower than the one reported in \citet{rodriguezdelpinoPropertiesionizedOutflows2019} where they detected ionized outflows through \Ha emission in $41\%$ of their AGN, although this discrepancy could be due to their significantly lower sample of AGN ($\sim30$) and a more relaxed criterion to identify outflows. Other works have found the fraction of outflows in AGN to be always larger than in sources with different type of ionization (star-formation, LIERs) and with detections rates varying between $10\%$ and $70\%$ as a function of AGN luminosity \citep{wooPrevalenceGasOutflows2016, averyIncidenceScalingRelations2021a,matzkoGalaxyPairsSloan2022}. Finally, given that our selection criteria (Section~\ref{subsec:AGN_selection}) might miss some Type 1 AGN and these generally show higher incidence than Type 2 \citep{pernaXraySDSSSample2017b, rojasBATAGNSpectroscopic2020}, the outflow fractions we obtain should be considered as lower limits. 

In the following sections we explore whether the incidence of outflows is affected by the environment where galaxies reside by studying its variation as a function of the environmental tracers described in Section~$\ref{subsec:environment_parameters}$.

\subsection{Isolated and satellite AGN}
\label{subsec:incidence_iso_sats}

We start by evaluating the fraction of ionized outflows in isolated and satellite AGN as a function of two main galaxy parameters: the stellar mass and the \OIII luminosity (\LOIII; estimated using the total flux in the \OIIIb line obtained in the spectral modelling described in Section~\ref{sec:analysis}). Controlling by stellar mass is important because environmental effects are expected to have different impact on galaxies depending on their stellar masses \citep{kuchnerEffectsClusterEnvironment2017, rodriguezdelpinoOMEGAOSIRISMapping2017, papovichEffectsEnvironmentEvolution2018} and the incidence of AGN in jellyfish galaxies are considered to be higher in more massive systems \citep{pelusoExploringAGNRamPressure2022}. Furthermore, since AGN accretion rate is known to drive powerful gas flows into the host \citep{kingPowerfulOutflowsFeedback2015a}, leading to the positive correlation between the incidence of outflows and AGN luminosity traced by \LOIII \citep[e.g.,][]{wylezalekionizedGasOutflow2020}, we also need to perform the comparison between isolated and satellite AGN accounting for possible differences in \LOIII between the two samples. Figure~$\ref{fig:Fraction_vs_mass_lum_cens_sats}$ contains the fraction of outflows in both AGN samples as a function of stellar mass (\emph{left}) and \LOIII (\emph{right}). The results presented in the left panel of this figure indicate that isolated AGN host a roughly uniform fraction of outflows ($\sim23\%$) throughout the stellar mass range covered in our study. In contrast, the AGN satellites display a clear trend of an increasing fraction of outflows towards more massive systems, ranging from $\sim7\%$ in the lower-mass bins to $32\%$ in galaxies with masses above $\sim10^{10.7}$M$_{\odot}$. With the exception of the highest-mass bin, the outflow incidence is generally higher in isolated AGN than in satellites, specially at masses below $\sim10^{10.3}$M$_{\odot}$ where the incidence is up to three times higher. At the highest mass bin the outflow incidence in satellites seems slightly higher than in isolated ones, although the fractions are consistent within the $1\sigma$ errors. With respect to the variation as a function of \LOIII, as shown in the right panel of Figure~$\ref{fig:Fraction_vs_mass_lum_cens_sats}$, the incidence of outflows in both AGN populations steeply increases from less than $5\%$ at the lowest luminosities ($<10^{39}$ergs$^{-1}$) to $>45\%$ in the more luminous systems ($>10^{41}$ergs$^{-1}$). Interestingly, within a given \LOIII bin the difference between the incidence of outflows in isolated and satellite AGN is not significant, with the fractions in both populations being consistent among them within the $1\sigma$ errors. 

The steeper increase in the incidence of outflows as a function of \LOIII together with the similar fractions obtained for isolated and satellite AGN within a given luminosity bin indicate that \LOIII is more strongly connected to the presence of ionized outflows than stellar mass. Given the more dominant correlation with \LOIII, the differences in the fraction of outflows observed at lower masses between isolated and satellite AGN might be related to a difference in \LOIII. We explore this possibility in Figure~\ref{fig:mass_lumOIII} by comparing the \LOIII values for isolated and satellite AGN in different stellar mass bins. At low stellar masses ($\lesssim10^{10.1}$M$_{\odot}$), the average luminosities of the two samples differ by $>0.3$dex, but this difference shrinks gradually towards higher stellar masses, with both populations displaying practically equal \LOIII distributions (mean values and standard deviation) in the more massive systems. These results demonstrate that the lower incidence of outflows in lower-mass satellite AGN is due to their lower \LOIII compared to isolated ones. Therefore, lower-mass satellite AGN must follow a different evolutionary path, undergoing processes that have a stronger influence at lower masses leading to a lower \LOIII{ } and, consequently, to a lower accretion rate, reducing the probability of hosting an ionized outflow. 

\begin{figure}
    \includegraphics[width=0.45\textwidth]{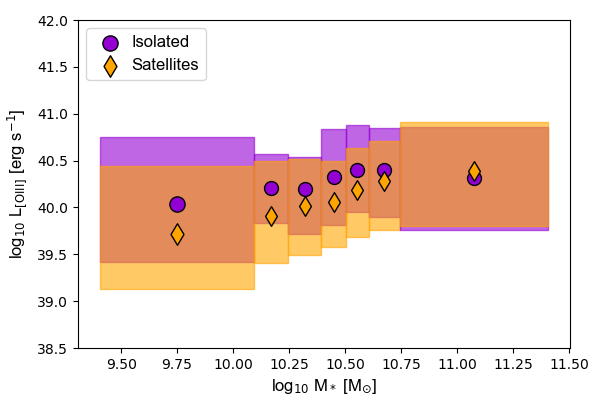}
    \caption[Logmass vs loglumOIII]{Average \LOIII in different stellar mass bins for isolated and satellite AGN. Bottom and top values of the boxes denote the 16th and 84th percentiles of the \LOIII distributions in each bin. Bins are defined to include at least 50 galaxies each.} 
    \label{fig:mass_lumOIII}
\end{figure}

To further explore the possible environmental influence on the incidence of outflows in satellite AGN we study its variation as a function of the normalized distance to the central galaxy of each group, $r_{\rm cen} / r_{\rm 180}$ (see Section~\ref{subsec:environment_parameters}). This is presented in the left panel of Figure~\ref{fig:Fraction_vs_distcentral_delta5} where the satellite population has been split into two samples by $10^{10.3}$M$_{\odot}$ (the stellar mass below which satellite AGN display lower incidence of outflows) to study the lower- and higher-mass regimes separately. The lower- and higher-mass satellite populations display similar trends with $r_{\rm cen} / r_{\rm 180}$, presenting lower fractions ($\sim5$\% and $\sim16-22$\%, respectively) at the closest distances to the central galaxy than at higher distances ($\sim13-18$\% and $\sim27-33$\%, respectively). As expected, the higher-mass sample always has higher fraction of outflows ($\geq0.1$dex) than the lower-mass ones. This result indicates that the processes responsible for the reduction in the incidence of outflows in satellite AGN should become more efficient in the proximity to the centre of the group or cluster.

\begin{figure*}
    \centering
    \begin{minipage}{.99\textwidth}
     \centering
        \includegraphics[width=0.48\textwidth]{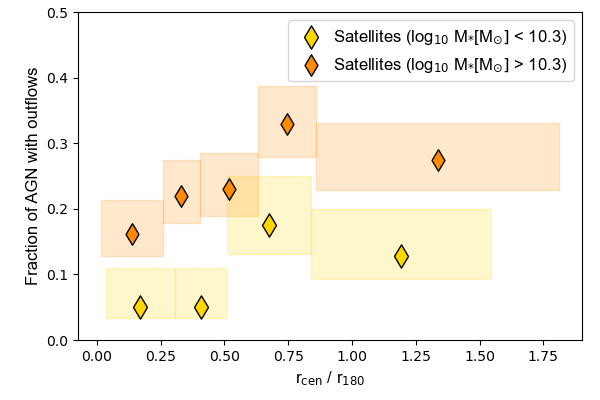}
        \includegraphics[width=0.48\textwidth]{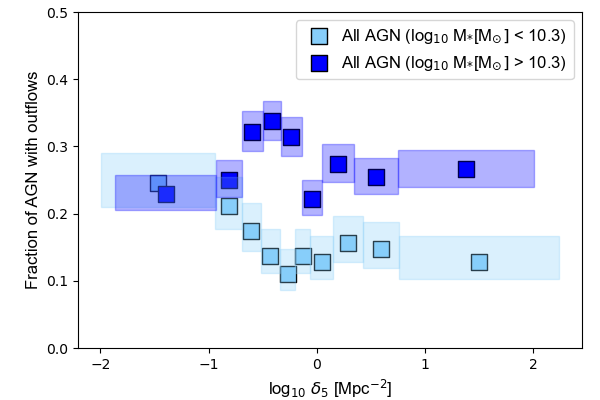}
        \caption[LOIII plots]{Fraction of AGN with outflows as a function of ($\emph{left}$) the normalized distance to the central galaxy of the group, $r_{\rm cen}$/$r_{\rm 180}$, for AGN satellites, and ($\emph{right}$) the projected surface density to the nearest 5th neighbour, $\delta_5$, for all the AGN in our sample, in both cases splitting the sample into lower- and higher-mass. Heights of the boxes denote the 1$\sigma$ errors associated to the fractions. Bins are defined to include at least 40 ($\emph{left}$) and 100 ($\emph{right}$) galaxies, respectively.} 
        \label{fig:Fraction_vs_distcentral_delta5}
    \end{minipage}
\end{figure*}

\subsection{Local surface density}
\label{subsection_incidence_surface_density}
Another way of exploring environmental effects is by probing the local environment around AGN, in our case through the projected surface density to the 5th neighbour, $\delta_5$ (see Section~\ref{subsec:environment_parameters}). A high value of $\delta_5$ might imply proximity between galaxies, favouring processes such as galaxy-galaxy harassment \citep{mooreGalaxyHarassmentEvolution1996}, and the presence of local inhomogeneities in the density of galaxies and/or associated intra-group/cumular material that could lead to ram-pressure stripping. Therefore, a high value of $\delta_5$ does not necessarily imply low $r_{\rm cen}$, since there can be groups of several (5 or more) galaxies being accreted by the cluster but still far from the centre, leading to a `pre-processing' effect \citep{fujitaPreProcessingGalaxiesEntering2004, darvishSpectroscopicStudyStarforming2015}. As in the previous section, to evaluate the influence of $\delta_5$ in the incidence of outflows we define lower- and higher-mass samples using $10^{10.3}$M$_{\odot}$ to divide, in this case, the full sample of AGN. The fraction of ionized outflows in the lower- and higher-mass samples of AGN as a function of local surface density are shown in the right panel of Figure \ref{fig:Fraction_vs_distcentral_delta5}. The two samples display clearly different trends: on the one hand, higher-mass AGN have a mean outflow incidence of $27\%$ across all densities, with some bins having fractions of $23\%$ (lowest $\delta_5$ bin) and $34\%$ (intermediate $\delta_5$ bins); on the other hand, the low-mass AGN experience a clear decline in the fraction of outflows towards denser regions, decreasing from $\sim25\%$ at low-densities to $\sim13\%$ at high densities. At these high densities, the fraction of outflows in higher-mass AGN is twice as high. The similar incidence of outflows at low densities and the diverging trends observed indicate that an increase in local density affects more strongly AGN with lower stellar masses, reducing significantly their probability of hosting ionized outflows.

\begin{figure*}
    \centering
    \begin{minipage}{.99\textwidth}
     \centering
        \includegraphics[width=0.48\textwidth]{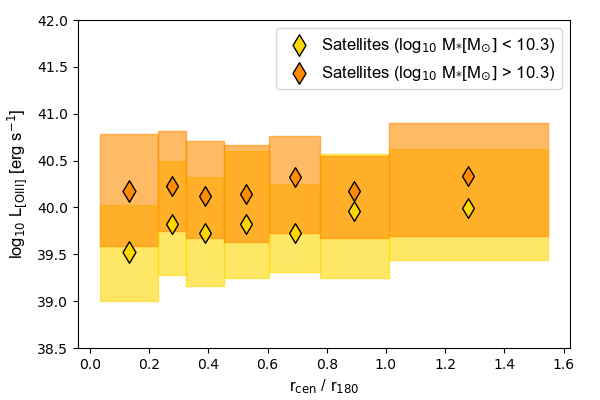}
        \includegraphics[width=0.48\textwidth]{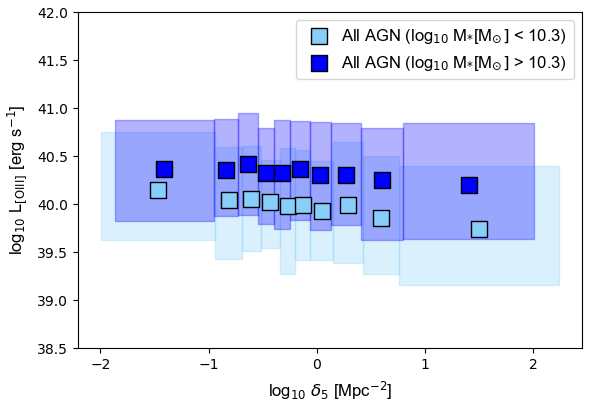}
        \caption[LOIII plots]{Average \LOIII in bins of ($\emph{left}$) normalized distance to the central galaxy of the group, $r_{\rm cen}$/$r_{\rm 180}$, for AGN satellites, and ($\emph{right}$) the projected surface density to the nearest 5th neighbour, $\delta_5$, for all the AGN in our sample, split into lower- and higher-mass samples. Bottom and top values of the boxes denote the 16th and 84th percentiles of the \LOIII distributions in each bin. Bins in each plot are defined to include at least 25 (\emph{left}) and 100 galaxies (\emph{right}) galaxies each.} 
        \label{fig:LOIII_plots}
    \end{minipage}
\end{figure*}

\section{Discussion}

\subsection{Environmental reduction of outflows in AGN}
The observed decline in the incidence of ionized outflows in low-mass satellite AGN (M$_{\odot}$~$<$~$10^{10.3}$M$_{\odot}$), together with the declining fractions at distances closer to the central galaxy of the group or cluster and towards locally denser regions indicate that environment plays a role in the likelihood of AGN hosting ionized outflows. Given that the triggering of AGN activity and ionized outflows depend on the availability of gas to act as fuel, it is reasonable to expect that the environmental processes responsible for a variation in the AGN fraction might also drive a change in the fractions of outflows in AGN. In this regard, previous works have associated the decline of AGN activity in clusters to two main drivers: a decrease in the merger rate, due to the higher relative velocities of galaxies in clusters \citep[i.e.,][]{popessoAGNFractionVelocity2006}, and to the stripping of gas from these systems by the dense intergalactic medium, preventing the feeding of the SMBH \citep{sabaterEffectInteractionsEnvironment2013, coldwellRelationSeyfertAccretion2014}. In support to the relevance of the merger rate as a driver, \citet{duplancicAGNsSmallGalaxy2021} found that the fraction of AGN in galaxy groups is always lower than in galaxy pairs and triplets, where AGN also tend to be more energetic. However, as noted by \citet{matzkoGalaxyPairsSloan2022}, for a given AGN luminosity galaxies undergoing mergers or interactions might not necessarily have a higher incidence of outflows than galaxies in isolation. Therefore, an increase in the incidence of outflows for interacting systems would only be expected if the interaction leads to an increase in luminosity. In our work we have removed potential merger candidates (see Section~\ref{subsec:environment_parameters}) to minimize their influence in the incidence of outflows in AGN and focus on environmental effects. However, if additional galaxy-galaxy interactions (not classified as mergers) that could trigger AGN activity and the subsequent outflow were present in our sample and favoured at lower densities \citep[e.g.,][]{alonsoGalaxyInteractionsII2012}, they could help explaining the decline in the incidence of outflows for the lower-mass AGN towards locally higher densities but not the roughly constant incidence in the higher-mass ones. 

In the case that galaxies experience gas stripping, there would be less fuel available to feed the SMBH and, as a consequence, the AGN luminosity, that is powered by accretion, should decrease. Since we have identified a decline in the fraction of outflows as a function of decreasing distance to the central galaxy and with increase local density (Figure~\ref{fig:Fraction_vs_distcentral_delta5}), we explore the variation of \LOIII as a function of these environmental parameters. This is done in Figure~\ref{fig:LOIII_plots} where we show the total \LOIII as a function of $r_{\rm cen}$/$r_{\rm 180}$, for AGN satellites, and $\delta_5$, for all AGN in our sample, in both cases splitting the samples into lower- and higher-mass galaxies by $10^{10.3}$M$_{\odot}$.

The left panel of this figure shows an increase in the \LOIII with $r_{\rm cen}$/$r_{\rm 180}$ for the lower-mass sample, whereas the higher-mass ones display relatively constant \LOIII distributions. A Spearman's correlation test for the lower-mass sample yields an associated coefficient $\rho=0.14$, with a p-value $<0.02$, although the scatter is quite large, as indicated by the 16th and 84th percentiles (height of the boxes in Figure~\ref{fig:LOIII_plots}). A similar decrease in total \LOIII for the lower-mass sample is observed towards regions of higher local density (right panel of the figure). In this case the correlation seems more significant, $\rho=-0.17$, p-value $<1\times10^{-6}$, although again the scatter is quite large. Moreover, as expected, the lower-mass AGN populations always have lower \LOIII than the higher-mass ones. 

These marginal, but systematic global trends observed in \LOIII could be an indication that environmental processes are affecting the gas reservoirs in galaxies, leading to a reduction in the available material to sustain the feeding of the AGN and subsequently reducing the incidence of ionized outflows. A potential candidate mechanism for causing this effect would be ram-pressure stripping, whose strength increases towards closer distances to the central galaxy, where relative velocities are higher, and at higher densities, only affecting the gas component in galaxies. 

The results we obtain are in apparent contrast with the positive connection between ram-pressure stripping and AGN activity suggested by \citet{poggiantiRampressureFeedingSupermassive2017} and \citet{pelusoExploringAGNRamPressure2022}, but in better agreement with the low fraction of jellyfish AGN found in other works such as \citet{roman-oliveiraOMEGAOSIRISMapping2019} and \cite{kolcuQuantifyingRoleRam2022}. However, in the comparison between these different works we have to bear in mind two key aspects: the environments and the range of stellar masses probed. For instance, although in this study we cover environments (see Figure~\ref{fig:Hist_halo_mass_sats}) where ram-pressure stripping has been found to be acting on galaxies leading to the jellyfish feature \citep{robertsLoTSSJellyfishGalaxies2021a}, SDSS data do no cover the more crowded regions of galaxy clusters \citep[e..g.,][]{gavazziCompleteCensusOptically2011} that probably host the most extreme jellyfish cases such as those studied in \citet{poggiantiRampressureFeedingSupermassive2017}. Moreover, in \citet{poggiantiRampressureFeedingSupermassive2017} they selected only very massive galaxies ($4\times10^{10}-3\times10^{11}$M$_{\odot}$) whereas in this work we consider AGN with a much wider range of stellar masses, specially at the lower-mass end. Therefore, the relative role of the environmental processes at different density regimes acting on galaxies with different stellar masses could explain the observed behaviours in these works.

\subsection{Impact on the regulation of metal content}

The important role that outflows are believed to play in the regulation of the metal content in galaxies implies that the lower fraction of ionized outflows in lower-mass satellite AGN found in this work might also alter the gas metallicity of these galaxies. In fact, a lower incidence of outflows will reduce the amount of gas that is expelled out of these lower-mass systems, allowing future star formation episodes than can enrich the metal content of their inter-stellar medium (ISM). As a consequence, these lower-mass satellite systems might not reduce their metallicities, contrary to those residing in lower-density environments (i.e., field) where the higher incidence of outflows will lead to their expected lower metallicities, following the well-known relation between stellar mass and gas metallicity \citep{tremontiOriginMassMetallicityRelation2004a, mannucciFundamentalRelationMass2010a, dayalPhysicsFundamentalMetallicity2013,chisholmMetalenrichedGalacticOutflows2018a}. 

Although an average increase in metallicity for galaxies residing in denser environments has been already reported in several works \citep{aracilHighmetallicityPhotoionizedGas2006, cooperRoleEnvironmentMassmetallicity2008, ellisonMassMetallicityRelation2009, darvishSpectroscopicStudyStarforming2015,schaeferSDSSIVMaNGAEvidence2019}, our findings that such differences are mainly relevant for the lower-mass galaxies are supported by the enhanced metal content of low-mass galaxies ($10^{8}-10^{10}$M$_{\odot}$) in the Coma and A1367 galaxy clusters reported in \citet{petropoulouENVIRONMENTALEFFECTSMETAL2012} and the 
results from \citet{wuDependenceMassmetallicityRelation2017} who reported that the difference in metallicity with environment in SDSS galaxies was only observed for the lower-mass ones, displaying a median metallicity systematically higher in denser regions. All these previous studies have associated the higher metallicities of galaxies in denser regions to processes such as the truncation of gas infall or wind re-accretion due to interactions with the ICM, and the accretion of metal-rich gas by satellite galaxies. The results obtained in this work indicate that, due to the lower incidence of ionized outflows in lower-mass AGN, their role as metal regulators might be less relevant, contributing to explain the higher metallicities of lower-mass galaxies in denser regions. 

\section{Conclusions}

In this work we have investigated whether environmental effects might have an impact on AGN activity by studying the incidence of ionized outflows in a sample of $\sim3300$ AGN residing in different environments drawn from SDSS DR13. Using the \OIIIb emission line as a tracer for the ionized gas we have searched for signatures of ionized outflows (found in $24\%$ of the AGN) and studied their incidence, comparing isolated and satellite AGN, and as a function of the projected distance to the central galaxy of the group and the projected surface density to the 5th neighbour. Our main results are: 

\begin{itemize}
\item At lower masses ($10^{9}-10^{10.3}$M$_{\odot}$), the fraction of ionized outflows is significantly lower in satellite than in isolated AGN. In satellites, the incidence increases strongly towards higher masses (from $\sim7$\% to 32\%), whereas in the isolated sample there is no significant variation within the stellar mass range we cover ($\sim10^9-10^{\rm11.5}$M$_{\odot}$). 
\\
\item The AGN luminosities, \LOIII, of isolated and satellites differ by $>0.3$dex at lower-masses, but this difference disappears gradually towards more massive systems. This lower \LOIII of satellites at lower masses can explain their lower fraction of outflows compared to isolated ones with similar stellar masses (previous point), since at fixed \LOIII the incidence is similar for both isolated and satellite AGN. 
\\
\item The fraction of lower-mass ($<$~$10^{10.3}$M$_{\odot}$) satellite AGN hosting ionized outflows declines significantly with projected surface density, from $\sim$~$25$\% to $\sim$~$13$\%. The fraction of outflows also decreases towards closer distances to the central galaxy of the group or cluster for all satellite AGN.  
\\
\item There is a large scatter in the relation between \LOIII and both the projected distance to the central and the projected surface density for satellite AGN. However, for the lower-mass ones there is a tentative indication of a decrease in \LOIII towards closer distances to the central and for denser regions.
\end{itemize}

Our findings indicate that environmental processes decrease the probability that AGN host ionized outflows, particularly less-massive systems ($10^{9}-10^{10.3}$M$_{\odot}$). We interpret this result as a consequence of the removal of the gas reservoirs around these galaxies due to the interaction with the dense intergalactic medium of group and cluster environments. This gas removal and lower incidence of outflows could also explain the lower fraction of AGN in clusters and the higher gas metallicities of galaxies in clusters compared to field ones, especially at lower masses, reported by previous works.

\begin{acknowledgements}
BRP, SA and MP acknowledge support from the Spanish Ministerio de Econom\'ia y Competitividad through the grants PID2019-106280GB-I00 and PID2021-127718NB-I00. ACS acknowledges funding from the Conselho Nacional de Desenvolvimento Científico e Tecnológico (CNPq) and the Rio Grande do Sul Research Foundation (FAPERGS) through grant CNPq-314301/2021-6 and FAPERGS/CAPES 19/2551-0000696-9. IL acknowledges support from the Spanish Ministry of Science and Innovation (MCIN) by means of the Recovery and Resilience Facility, and the Agencia Estatal de Investigación (AEI) under the projects BDC20221289 and PID2019-105423GA-I00. MP acknowledges support from the Programa Atracci\'on de Talento de la Comunidad de Madrid via grant 2018-T2/TIC-11715. Funding for the Sloan Digital Sky Survey IV has been provided by the Alfred P. Sloan Foundation, the U.S. Department of Energy Office of Science, and the Participating Institutions. SDSS-IV acknowledges support and resources from the centre for High-Performance Computing at the University of Utah. The SDSS web site is www.sdss.org. This research made use of Astropy, a community-developed core Python package for Astronomy \citep{astropycollaborationAstropyProjectBuilding2018}. This research has made use of the NASA/IPAC Extragalactic Database (NED) which is operated by the Jet Propulsion Laboratory, California Institute of Technology, under contract with the National Aeronautics and Space Administration. 
\end{acknowledgements}

%
%
\bibliographystyle{aa}
\bibliography{refs}

\end{document}